\documentclass[epj,numbook]{svjour}
\usepackage{graphics,epsfig,amsmath,amssymb}
\begin{document}
\title{Opening of a gap in  an inhomogeneous external field.}
\titlerunning{Gap's in the antiferromagnetic Heisenberg model}
\authorrunning{Fledderjohann}
\author{A. Fledderjohann, M. Karbach and K.-H. M\"utter}
\institute{Physics Department, University of Wuppertal, D-42097 Wuppertal, Germany}
\date{\today}
\abstract{We study the one-dimensional spin-1/2 antiferromagnetic Heisenberg
  model exposed to an external field, which is a superposition of a homogeneous
  field $h_{3}$ and a small periodic field of strength $h_{1}$. For the case of
  a transverse staggered field a gap opens, which scales with
  $h_{1}^{\epsilon_{1}}$, where $\epsilon_{1}=\epsilon_{1}(h_{3})$ is given by
  the critical exponent $\eta_{1}(M(h_{3}))$ defined through the transverse
  structure factor of the model at $h_{1}=0$. For the case of a longitudinal
  periodic field with wave vector $q=\pi/2$ and strength $h_{q}$ a plateau is
  found in the magnetization curve at $M=1/4$. The difference of the upper- and
  lower magnetic field scales with $h_{3}^{u}-h_{3}^{l}\sim
  h_{q}^{\epsilon_{3}}$, where $\epsilon_{3}=\epsilon_{3}(h_{3})$ is given by
  the critical exponent $\eta_{3}(M(h_{3}))$ defined through the longitudinal
  structure factor of the model at $h_{q}=0$.}

\PACS{{75.10 -b}{General theory and models of magnetic ordering}} 
\maketitle
%
\section{Introduction}
\label{sec:Introduction}
%
The properties of the one-dimensional (1D) spin-1/2 antiferromagnetic Heisenberg
model (AFH) with nearest neighbour coupling:
\begin{eqnarray}
  \label{eq:1}
  {\bf H}(h_{3}) &\equiv& {\bf H}_{0}-2h_{3}{\bf S}_3(0), \\
  {\bf H}_{0} &\equiv& 2\sum_{l=1}^{N}{\bf S}_l \cdot {\bf S}_{l+1}, \\
  {\bf S}_a(q) &\equiv& \sum_{l=1}^{N}e^{ilq}S_l^{a},\quad a=1,2,3,
\end{eqnarray}
in the presence of a homogeneous external field of strength $h_{3}$ are well known:
\begin{enumerate}
\item There is no gap. The magnetization curve $M=M(h_{3})$ is a monotonically
increasing convex function \cite{BF64,Grif64,YY66c} for $h_{3}\geq 0$; in
particular there is no plateau.
\item In the presence of the field $h_{3}$ the ground state $|p_{s},S\rangle$ of
  ${\bf H}(h_{3})$ has total spin $S=S^{3}_{T}=NM(h_3)$ and momentum $p_{s}=0,\pi$
  -- depending on $S$ and $N$.
\item The low energy excitations which can be reached from the ground state
  $|p_{s},S\rangle$ by means of the transition operators ${\bf S}_3(q)$ and ${\bf
    S}_\pm(q)$: 
  \begin{eqnarray}
    \label{eq:2}
    \omega_{3}(q,h_{3}) &=& E(p_{s}+q,S)-E(p_{s},S), \\
    \label{eq:2b}
    \omega_{\pm}(q,h_{3}) &=& \nonumber 
    E(p_{s}+q,S\pm 1)-E(p_{s},S)\pm h_{3}, \nonumber \\
  \end{eqnarray}
  vanish at the \textit{soft mode} momenta $q_{a}=q_{a}(M)$:
  \begin{equation}
    \label{eq:3}
    \hat\Omega_{a}(M) \equiv \lim_{N\to\infty} N\omega_{a}(q_{a}(M),h_{3}),
  \end{equation}
  with
  \begin{equation}
    \label{eq:4}
    q_{a}(M) = \pi
    \begin{cases}
      1 &: a=1,2 \\ 1-2M &: a=3 
    \end{cases}.
  \end{equation}
  Conformal field theory describes the critical behaviour at the soft modes
  \cite{BIK86,BIR87,MZ91,CH93,FGM+96}. In particular the field dependence of the
  $\eta$-exponents:
  \begin{equation}
    \label{eq:5}
    \eta_{a}(M)=\frac{\hat\Omega_{a}(M)}{\pi v(M)}
  \end{equation}
  has been computed by means of the Bethe Ansatz solutions for the energy
  differences and the spin wave velocity \cite{FMK98a,Klue98a}
  \begin{equation}
    \label{eq:6}
    v(M) \!=\! \frac{1}{2\pi} \!\lim_{N\to\infty}\!
    N[ E(p_{s}\!+\!2\pi/N,S)\!-\!E(p_{s},S)]
  \end{equation}
\item The $\eta$-exponents govern the finite-size behaviour of the transition
  amplitudes:
  \begin{eqnarray}
    \label{eq:7a}
    \langle S\pm 1,p_{s}+\pi | {\bf S}_{\pm}(\pi) | S,p_{s}\rangle 
    &\stackrel{N\rightarrow\infty}{\longrightarrow}& N^{\kappa_{1}(h_{3})} \\
    \label{eq:7b}
    \langle S,p_{s}\!+\!q_{3}|{\bf S}_{3}(q_{3}) | S,p_{s}\rangle 
    &\stackrel{N\rightarrow\infty}{\longrightarrow}& N^{\kappa_{3}(h_{3})}
  \end{eqnarray}
  with
  \begin{equation}
    \label{eq:8}
    \kappa_{a}(h_{3}) = 1-\frac{\eta_{a}(M(h_{3}))}{2}, 
  \end{equation}
and of the static structure factors:
\begin{equation}
  \label{eq:9}
  \langle S,p_{s}|{\bf S}_{a}(q_{a}){\bf S}_{a}(q_{a})^{\dagger}|S,p_{s}\rangle
  \stackrel{N\rightarrow\infty}{\longrightarrow} 
  N^{2-\eta_{a}(M)}.
\end{equation}
At the soft mode momenta $q_{a}=q_{a}(M)$ the dynamical structure factors develop
infrared singularities of the type $\omega^{-2\kappa_{a}(h_{3})}$.
\end{enumerate}
First evidence for the existence of \textit{low energy modes} in the excitation
spectrum has been found recently in neutron scattering experiments on copper
benzoat, \cite{DDR+96,DHR+97} exposed to a homogeneous magnetic field $h_{3}$.
An exponential fit to the temperature dependence of the specific heat data
revealed, however, that there is a gap in the energy
differences~(\ref{eq:2}),(\ref{eq:2b}) and~(\ref{eq:4}) , which opens with the
field strength $h_{3}$ as $h_{3}^{\epsilon},\;\epsilon=2/3$. This means of
course, that the compound copper benzoat can not be described by a 1D Heisenberg
antiferromagnet. Oshikawa and Affleck \cite{OA97} argued that the local
$g$-tensor for the Cu ions generates an effective staggered field of strength
($h_{1}\ll h_{3}$), perpendicular to the uniform field $h_{3}$. Therefore, one
is lead to investigate the Hamiltonian:
 \begin{equation}
   \label{eq:11}
  {\bf H}(h_{3},h_{1}) \equiv {\bf H}(h_{3})+ 2h_{1}{\bf S}_1(\pi).   
 \end{equation}
It is the purpose of this paper to study the evolution of the gaps 
\begin{equation}
  \label{eq:12}
  \omega_{a}(q_{a},h_{3},h_{1}) \propto h_{1}^{\epsilon_{a}(h_{3})},
\end{equation}
by switching on the transverse staggered field.  In particular we are interested
in the $h_{3}$-dependence of the exponents $\epsilon_{a}(h_{3})$.

It has been pointed out by the authors of Ref. \cite{OA97} that a staggered
field alone, i.e. $h_{3}=0, M=0$, generates a ground state gap which opens with
$h_{1}^{\epsilon},\;\epsilon=2/3 $.  In a previous paper we have studied the
finite-size scaling behaviour of the gap and of the staggered magnetization in
the scaling limit $h_{1}\to 0,\;N\to\infty$ and fixed scaling variable
$x=Nh_{1}^{\epsilon}$ at $M=0$.

The method used in Ref. \cite{FMK98a} is based on a closed set of differential
equations, which describes the $h_{1}$-evolution of the energy gap
$\omega_{3}(\pi,0,h_{1})$ [Eq.~(\ref{eq:2})] and of the relevant transition
amplitude~(\ref{eq:7b}) for $h_{3}=0$. It turns out that the exponent
$\epsilon(h_{3}=0)$ in~~(\ref{eq:12}) is fixed by the finite-size behaviour of the
initial values, i.e.~(\ref{eq:2}) and~(\ref{eq:7b}) for $h_{3}=h_{1}=0$:
\begin{equation}
  \label{eq:14}
  \epsilon(h_{3}=0) = \frac{1}{1+\kappa(0)} = \frac{2}{3}. 
\end{equation}
In this paper, we extend the method of Ref.\cite{FMK98a} to the case $h_{3}>0$. 

In section \ref{sec:Evolution-equations} we discuss the evolution equations for
the Hamiltonian~(\ref{eq:11}). The finite-size behaviour of the initial
conditions ($h_{1}=0,h_{3}>0$) for the gaps~(\ref{eq:2}) and (\ref{eq:2b}) and
for the relevant transition matrix elements~(\ref{eq:7a}) and (\ref{eq:7b}) is
reviewed as well.

Switching on the transverse staggered field in~(\ref{eq:11}) a gap opens at the
field independent [Eq.~(\ref{eq:2b}) for $q=\pi$] and the field dependent
[Eq.~(\ref{eq:2}) for $q=q_{3}(M)$] soft modes. The finite-size scaling behaviour of
these gaps is studied in sections~\ref{sec:gap-ind-soft mode}
and~\ref{sec:gap-dep-soft mode}, respectively. In
section~\ref{sec:gap-longitudinal}, we investigate the effect of a longitudinal
periodic field on the low-energy excitations of the AFH model. From these
results we infer in section~\ref{sec:mag} the corresponding magnetization curve.
%
\section{Evolution equation and initial conditions}
\label{sec:Evolution-equations}
%
Starting from the eigenvalue equation of the Hamiltonian~(\ref{eq:11}) 
\begin{eqnarray}
  \label{eq:HPsi}
  {\bf H}(h_{3},h_{1})|\Psi_n(h_{3},h_{1})
    \rangle &=& E_n(h_{3},h_{1})|\Psi_n(h_3,h_1)\rangle,
\end{eqnarray}
it is straight forward to derive the following set of differential equations
\begin{eqnarray}
  \frac{d^{2} E_n}{dh_{1}^{2}} &=&  
   -2\sum_{l\neq n}\frac{|T_{ln}|^{2}}{\omega_{ln}}, \label{eq:d2En}\\
  \frac{d T_{nm}}{dh_{1}} &=&  
   \!-\!\sum_{l\neq m,n}\left[
    \frac{T_{nl}T_{lm}}{\omega_{ln}} + \frac{T_{nl}T_{lm}}{\omega_{lm}}
  \right]
  -\frac{T_{nm}}{\omega_{nm}}\frac{d\omega_{nm}}{dh_{1}}, \label{eq:dSmn}
 \nonumber \\
\end{eqnarray}
which describes the evolution of the energy eigenvalues $E_{n}=E_{n}(h_3,h_1)$,
energy differences $\omega_{nm}=\omega_{nm}(h_3,h_1)=E_{n}-E_{m}$ and transition
matrix elements
\begin{equation}
  \label{eq:15}
 T_{nm}(h_{3},h_{1}) \equiv 
 \langle \Psi_n(h_3,h_1)|{\bf S}_{1}(\pi)|\Psi_m(h_3,h_1) \rangle,  
\end{equation}
of the perturbation operator ${\bf S}_{1}(\pi)$.\footnote{The $N$-dependence of
  eigenvalues and transition matrix elements is always understood.} The latter
has the following properties: It changes the momentum by $\Delta p= \pi$ and the
total spin $S_{T}^{3}$ by one unit. Therefore, the eigenstates
$|\Psi_{n}(h_3,h_1)\rangle$ are linear combinations
\begin{eqnarray}
  \label{eq:16}
  |\Psi_{n}(h_3,h_1)\rangle &=& \sum_{S_{T}^{3}} \left[
  a_{n}(S_{T}^{3},h_{1})|p_{n},S_{T}^{3}\rangle \right. \nonumber \\ &&
    ~~~ \left. + b_{n}(S_{T}^{3},h_{1})|p_{n}\!+\!\pi,S_{T}^{3}\rangle\right],
\end{eqnarray}
of eigenstates $|p_{n},S_{T}^{3}\rangle$ and $|p_{n}\!+\!\pi,S_{T}^{3}\rangle$
to the total spin $S_{T}^{3}$ and the momenta $p_{n},p_{n}+\pi$. Note, that the
evolution equations~(\ref{eq:d2En}) and~(\ref{eq:dSmn}) decouple for different
momenta $p_{n},p_{m}$ with $|p_{n}-p_{m}| \neq \pi$. In
section~\ref{sec:gap-ind-soft mode} and~\ref{sec:gap-dep-soft mode} we will
study the following cases:
\begin{center}
\begin{enumerate}
\item $p_{n}=0,\pi$,
\item $p_{n}=q_{3}(M),\;q_{3}(M)+\pi$.
\end{enumerate}
\end{center}
For both cases we have the  initial conditions:
\begin{eqnarray}
  \label{eq:23a}
  \omega_{nm}(q,h_3,h_{1}=0) &=& \frac{a_{nm}(h_3)}{N}, \\ \label{eq:23b}
  T_{nm}(h_{3},h_{1}=0) &=& b_{nm}(h_3)N^{\kappa(h_3)},
\end{eqnarray}
which are completely fixed by the excitation energies and transition amplitudes
of the unperturbed problem $(h_{1}=0)$ in a uniform field $h_{3}$. We can now
repeat the whole line of arguments, we developed for $h_{3}=0$ in
Ref.~\cite{FMK98a}. The evolution equations~(\ref{eq:d2En}) and~(\ref{eq:dSmn})
possess scaling solutions:
\begin{eqnarray}
  \label{eq:29a}
    \omega_{nm}(q,h_3,h_1) &=& h_{1}^{\epsilon(h_3)}\Omega_{nm}(x), 
\\ \label{eq:29b}
         T_{nm}(h_3,h_1) &=& Nh_{1}^{\sigma(h_3)}\Theta_{nm}(x),
\end{eqnarray}
in the combined limit
\begin{equation}
  \label{eq:30}
  h_{1}\to 0,\quad N\to\infty,\quad x\equiv Nh_{1}^{\epsilon(h_3)}\; \mbox{fixed}.
\end{equation}
The exponents $\epsilon(h_3)$ and $\sigma(h_3)$ are given by the finite-size
behaviour of the initial values~(\ref{eq:23a}) and (\ref{eq:23b}):
\begin{equation}
  \label{eq:7}
  \epsilon(h_3) = \frac{1}{1+\kappa(h_3)}, \quad
  \sigma(h_3) = \frac{1-\kappa(h_3)}{1+\kappa(h_3)}.
\end{equation}
%
\subsection{The gap at the field independent soft mode $q=\pi$}
\label{sec:gap-ind-soft mode}
\begin{figure}[ht]
\centerline{\epsfig{file=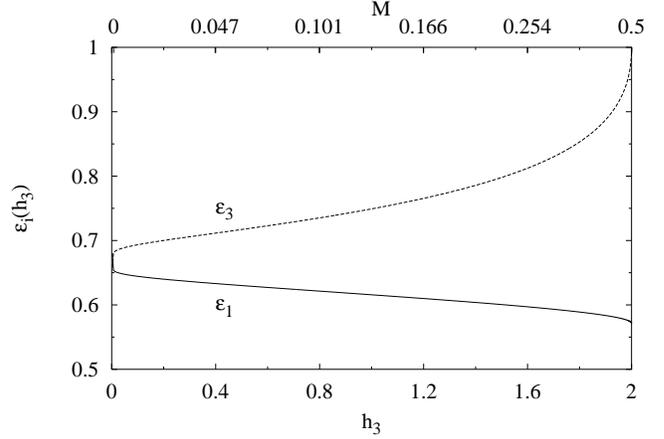,width=6.0cm,angle=-90}}
\caption{The exact critical exponents $\epsilon_{1}$ (solid line) and
  $\epsilon_{3}$ (dashed line) versus $h_{3}$ and $M$, determined from a Bethe
  ansatz solution of finite system size $N=4096$.}
\label{fig:eta}
\end{figure}

\begin{figure}[ht]
\centerline{\hspace*{4cm}\epsfig{file=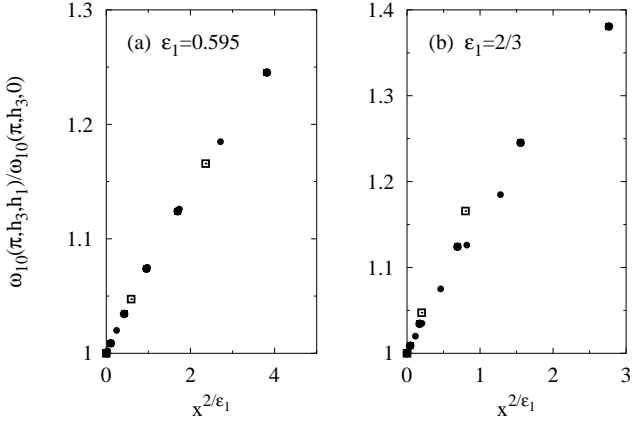,width=6.0cm,angle=-90}}
\caption{A comparison of the ratio~(\ref{eq:31}) for two different values of
  $\epsilon_{1}$ and the midpoint field $\bar h_{3}$ [Eq.~(\ref{eq:35})] for
  system sizes $N=8,12,16,20$}
\label{fig:fig:e10xB}
\end{figure}

%
As was pointed out in the introduction, the ground state
$|n=0\rangle=|p_{s},S\rangle$ of the 1D spin-1/2 AFH model, ${\bf H}(h_{3},0)$,
in the presence of a uniform field $h_{3}$ has total spin $S_{T}^{3}=S=NM(h_3)$
and momentum $p_{s}=0$, or $p_{s}=\pi$. The first excited state which can be
reached with the operator ${\bf S}_{1}(\pi)$:
\begin{equation}
  \label{eq:20}
  |n=\pm 1\rangle = |p_{s}+\pi,S_{T}^{3}=S\pm 1\rangle,
\end{equation}
has a gap of the type~(\ref{eq:23a})
\begin{equation}
  \label{eq:21}
  \omega_{\pm 10}(\pi,h_3,0)=E(p_{s}+\pi,S\pm 1)-E(p_{s},S) \mp h_{3},
\end{equation}
which vanishes as $N^{-1}$ for $N\to\infty$. The transition matrix elements:
\begin{equation}
  \label{eq:22}
  T_{\pm 10}(h_3,0) \equiv \langle \pm 1 | {\bf S}_{\pm}(\pi) | 0\rangle 
  \stackrel{N\to\infty}{\longrightarrow} N^{\kappa_{1}(h_3)}
\end{equation}
diverge in the limit $N\to\infty$, where $\kappa_{1}(h_3)$ is obtained from the
known $\eta_{1}(M)$ exponent~(\ref{eq:8}).  Both curves, $\eta_{1}=\eta_{1}(M)$
and $M=M(h_3)$ were determined exactly by means of Bethe ansatz solutions on
large systems \cite{FGM+96}, as well via a solution of a system of nonlinear
integral equation derived from the Bethe Ansatz~\cite{Klue98a}. The
$h_{3}$-dependence is shown in Fig.~\ref{fig:eta}. It starts at the known value
$\epsilon_{1}(h_3=0)=2/3$ and then drops monotonically with $h_3$. At
$h_{3}(M=1/4)=1.58\ldots$, the exponent is reduced to
\begin{equation}
  \label{eq:epsilon-m-1d4}
  \epsilon_{1}(h_3(1/4))=0.5975\ldots.
\end{equation}
In order to explore the scaling behaviour~(\ref{eq:29a}) of the gap, we have
determined numerically the ratios:
\begin{eqnarray}
  \label{eq:31}
  \frac{\omega_{10}(\pi,h_3,h_1)}{\omega_{10}(\pi,h_{3},0)} 
            &=& 1 + e_{10}(x,h_3) \nonumber, \\
             & = & 1+x\frac{\Omega_{10}(x,h_3)}{a_{10}},
\end{eqnarray}
with $x=Nh_{1}^{\epsilon_{1}(h_{3})}$ and $\Omega_{10}$ as given in
Ref.~\cite{FMK98a}, on finite systems. The homogeneous field $h_3$ has to be
chosen carefully.  According to our premise, the ground state
$|0\rangle=|p_{s},S\rangle$ (at $h_{1}=0$) has total spin $S_{T}^{3}=S=NM(h_3)$
and energy $E(p_{s},S)-2h_{3}S_{T}^{z}$. The two excited states $|\pm 1\rangle =
|p_{s}+\pi,S_{T}^{3}=S\pm 1\rangle$ have a gap. Positivity of the gaps yields an
upper and lower bound of the $h_{3}$-field ($h_{3}^{u}\geq h_{3}\geq
h_{3}^{l}$):
\begin{eqnarray}
  \label{eq:33}
  2h_3^u &=& E(p_{s}\!+\!\pi,S\!+\!1) - E(p_{s},S), 
  \\ \label{eq:33b}
  2h_3^l &=& E(p_{s},S) - E(p_{s}\!+\!\pi,S\!-\!1),
\end{eqnarray}
which leads to the well known steps in the magnetization curve on finite
systems\cite{BF64}.  Note, that at the edges $h_3^u$ and $h_3^l$ the excitations
energies:
\begin{eqnarray}
  \label{eq:34}
  \omega_{+10}(\pi,h_3^u,h_{1}=0) &=& 0, \\
  \omega_{-10}(\pi,h_3^l,h_{1}=0) &=& 0, 
\end{eqnarray}
vanish identically. Therefore, ratios of the gap~(\ref{eq:31}) do not make
sense in these cases. At the midpoint field $\bar{h}_{3}$, however:
\begin{eqnarray}
  \label{eq:35}
  2\bar{h}_{3} &\equiv& (h_3^{u}+h_3^l)/2 \nonumber \\
          &=& [E(p_{s}\!+\!\pi,S\!+\!1)-E(p_{s}\!+\!\pi,S\!-\!1)]/2,
\end{eqnarray}
the two excited states have the same gap:
\begin{eqnarray}
  \label{eq:36}
  \omega_{\pm 10}(\pi,\bar h_{3},0) &=& \nonumber \\
     && \hspace{-20mm} =\frac{E(p_{s}\!+\!\pi,S\!+\!1)+E(p_{s}\!+\!\pi,S\!-\!1)-2
               E(p_{s},S)}{2}. \nonumber \\
\end{eqnarray}
The degeneracy of these two excited states is not lifted in the first oder
perturbation theory in $h_{1}$, since all the relevant matrix elements
\begin{equation}
  \label{eq:37}
 \langle n| {\bf S}_{1}(\pi) | m\rangle = 0,\quad n,m=\pm 1
\end{equation}
vanish. The ratio~(\ref{eq:31}) is shown in Fig.~\ref{fig:fig:e10xB}(a), for the
midpoint field $\bar h_{3}=\bar h_{3}(N)\approx1.58$, corresponding to a
magnetization $M=1/4$ on system sizes $N=8,12,16,20$. Optimal scaling is
achieved here, with the exponent $\epsilon_{1}=0.595(5)$, which is in excellent
agreement with the exact value~(\ref{eq:epsilon-m-1d4}).
According to Ref.~\cite{FMK98a}, the low $x$-behaviour of the scaling function
$e_{10}(x,h_{3})$ is also predicted by the evolution equations~(\ref{eq:d2En})
and~(\ref{eq:dSmn}) in the scaling limit~(\ref{eq:30}):
\begin{equation}
  \label{eq:38}
  e_{10}(x,h_3) = e_{10}(h_3)x^{\phi_{1}(h_3)},
\end{equation}
with $\phi_{1}(h_3)=2/\epsilon_{1}(h_3)$. The linear behaviour in the variable
$x^{2/\epsilon_{1}(h_3)}$ for small $x$-values is clearly seen in
Fig.~\ref{fig:fig:e10xB}(a).

The effect of the homogeneous $h_3$-field on the exponent $\epsilon_{1}$ is
demonstrated in Fig.~\ref{fig:fig:e10xB}(a). An exponent
$\epsilon_{1}(h_3)=\epsilon_{1}(h_3=0)=2/3$ independent of $h_3$ would lead to
considerable scaling violations of the ratios~(\ref{eq:31}), as is demonstrated
in Fig.~\ref{fig:fig:e10xB}(b).
%
\subsection{The gap at the field dependent soft mode $q=q_{3}$}
\label{sec:gap-dep-soft mode}
\begin{figure}[ht]
\centerline{\epsfig{file=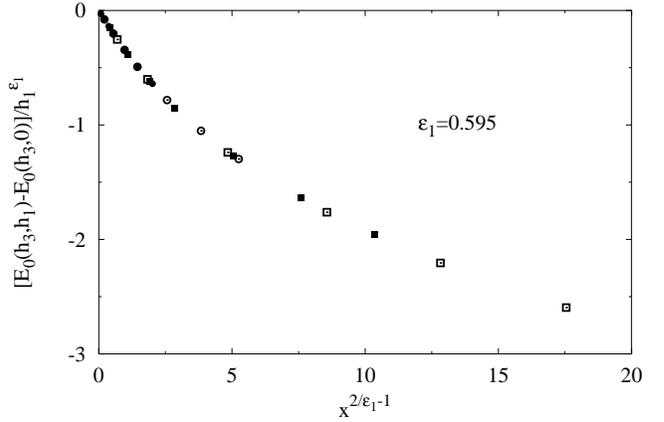,width=6.0cm,angle=-90}}
\caption{The scaling of the ground state energy~(\ref{eq:24}) for the midpoint
  field $\bar h_{3}$ [Eq.~(\ref{eq:23})] in the limit~(\ref{eq:30}). Numerical
  data were obtained on system sizes $N=8,12,16,20$. }
\label{fig:fig_e0-eps=0.59}
\end{figure}

%
Let us now turn to the field dependent soft mode [Eq.~(\ref{eq:2}) for
$q=q_{3}(M)$]. Switching on the perturbation operator $h_{1}{\bf S}_{1}(\pi)$
the ground state energy $E(p_{s},S)$ and the energy $E(p_{s}\!+\!q_{3}(M),S)$ of
the excited state evolve independently, since their momentum difference
$q_{3}(M)$ does not fit to the momentum transfer $\pi$ mediated by the operator
${\bf S}_{1}(\pi)$. In other words, we have to study the ground state energy
$E_{0}(h_3,h_1)$ in the sectors with momentum $p=0,\pi$ and
$p=q_{3}(M),q_{3}(M)+\pi$, separately. In both cases insertion of the scaling
ansatz~(\ref{eq:29a}) and (\ref{eq:29b}) for the excitation energies
$\omega_{n0}$ and transition amplitudes $T_{n0}$ into~(\ref{eq:d2En}) yields:
\begin{equation}
  \label{eq:17}
  \frac{d^{2}E_{0}}{dh_{1}^{2}} = -N^{1+2\kappa_{1}(h_3)}x^{1-2\kappa_{1}(h_3)}
  \sum_{l\neq 0}\frac{|\Theta_{l0}(x)|^{2}}{\Omega_{l0}(x)},
\end{equation}
where $x=Nh_{1}^{\epsilon_{1}(h_3)}$. To integrate~(\ref{eq:17}) we introduce 
\begin{equation}
  \label{eq:18}
  y\equiv x^{1/\epsilon_{1}(h_3)} = h_{1}N^{1+\kappa_{1}(h_3)},
\end{equation}
and
\begin{equation}
  \label{eq:19}
  f(y)\equiv x^{1-2\kappa_{1}(h_3)}
       \sum_{l\neq 0}\frac{|\Theta_{l0}(x)|^{2}}{\Omega_{l0}(x)},
\end{equation}
from which follows:
\begin{eqnarray}
  \label{eq:23}
  E_{0}(h_3,h_1)-E_{0}(h_3,0) &=& \nonumber\\
   && \hspace{-3cm} 
   -\left(\frac{h_{1}}{y}\right)^{\epsilon_{1}(h_3)} 
    \int_{0}^{y}dy'\int_{0}^{y'}dy'' f(y''). 
\end{eqnarray}
Here, we have used the fact, that 
\begin{equation}
  \label{eq:24}
  \left.\frac{dE_{0}(h_3,h_1)}{dh_{1}}\right|_{h_{1}=0} = 
   \langle 0|{\bf S}_{1}(\pi) | 0 \rangle|_{h_{1}=0} =0.
\end{equation}
Equation~(\ref{eq:23}) describes the lowering of the ground state energy, if we
switch on the staggered field of strength $h_{1}$. We observe the same scaling
behaviour with $h_{1}^{\epsilon_{1}}$, we found for the excitation energies
$\omega_{n0}(\pi,h_{3},h_{1})$. In Fig.~\ref{fig:fig_e0-eps=0.59} we have plotted
$[E_{0}(h_3,h_1)-E_{0}(h_{3},0)]/h_{1}^{\epsilon_{1}(h_3)}$ versus the scaling
variable $x^{2/\epsilon_{1}-1}$ for the case $p=0,\pi$.
We observe a linear behaviour in this variable, which is a consequence of the
small $x$-behaviour of the energy differences $\Omega_{l0}(x)$ and transition
amplitudes $\Theta_{l0}(x)$ in Eq.~(\ref{eq:19}) [See Ref.~\cite{FMK98a}]:
\begin{equation}
  \label{eq:10}
  \Omega_{l0}(x) \sim \frac{a_{l0}}{x}, \quad 
  \Theta_{l0}(x)  \sim x^{-2+1/\epsilon_{1}}.
\end{equation}
Therefore, the integrand on the right-hand side of~(\ref{eq:23}) is constant and
the small $x$-behaviour of~(\ref{eq:23}) is governed by
$y^{2-\epsilon_{1}}=x^{2/\epsilon_{1}-1}$.

Let us next turn to the lowering of the ground state energy in the sector with
$p=q_{3}(M),q_{3}(M)+\pi$. The exponents $\kappa_{\pm}(h_3)$ are defined by the
initial conditions $(h_{1}=0)$ for the transition matrix elements:
\begin{eqnarray}
  \label{eq:25}
  \langle\pm 1 | {\bf S}_{1}(\pi) | 0 \rangle &=& 
  \langle p_{s\pm 1}\!+\!q_{3}(M),S\!\pm \!1 |  {\bf S}_{1}(\pi) | 
           p_{s}\!+\!q_{3}(M),S \rangle \nonumber \\
           &=& b_{\pm 10}(h_{3}) N^{\kappa_{\pm}(h_3)}.
\end{eqnarray}
Conformal field theory relates the corresponding $\eta$-expo\-nents
($\kappa_{\pm} = 1-\eta_{\pm}/2$) to the scaled energy differences:
\begin{equation}
  \label{eq:26}
  \eta_{\pm}(M) = \frac{\hat\Omega_{\pm}(M)}{\pi v(M)},
\end{equation}
with
\begin{eqnarray}
  \label{eq:27a}
  \hat\Omega_{\pm}(M) &=& \lim_{N\to\infty} N 
  \left[E(p_{s\pm 1}\!+\!q_{3}(M),S\!\pm\! 1)\right. \nonumber \\ &&\hspace{1cm}
        \left. -E(p_{s}\!+\!q_{3}(M),S)\right].
\end{eqnarray}
Here $v(M)$ is the spin wave velocity~(\ref{eq:6}) at the soft mode $q=0$.
Evaluating~(\ref{eq:27a}) and~(\ref{eq:6}) leads to the following
representation of the $\eta_{\pm}$-exponents~(\ref{eq:26})
\begin{equation}
  \label{eq:27}
  \eta_{\pm}(M) = \eta_{1}(M) + \frac{v_{\pm}(M)}{v(M)},
\end{equation}
where
\begin{eqnarray}
   \label{eq:vpm} 
  v_{\pm}(M)
    &=&  \frac{1}{2\pi} \!\lim_{N\to\infty}\!
        N[ E(p_{s\pm 1}\!+\!q_{3}(M\!\pm\!1/N)\!\pm\!2\pi/N,S\!\pm\! 1)\nonumber
\\ && \hspace{1cm} - E(p_{s\pm 1}+q_{3}(M\!\pm\!1/N),S\pm 1)], 
\end{eqnarray}
are the right-hand- (+) and left-hand (-) spin wave velocities obtained from the
slopes of the dispersion curve  approaching the soft mode momentum
from the right- and from the left-hand side, respectively:
\begin{equation}
  \label{eq:28}
  p \mapsto p_{s}+q_{3}(M) \pm 2\pi/N.
\end{equation}
From conformal invariance arguments for the energy differences in~(\ref{eq:vpm})
we get
\begin{equation}
  \label{eq:13}
  \eta_{+}(M)=\eta_{-}(M) = 1+\eta_{1}(M).
\end{equation}
In summary, we conclude that the gap of the field dependent soft mode $q_{3}(M)$: 
\begin{equation}
  \label{eq:29}
  E(p_{s}\!+\!q_{3}(M),S)-E(p_{s},S) \sim h_{1}^{\epsilon_{1}(h_{3})},
\end{equation}
is dominated by the lowering of ground state energy $E(p_{s},S)$ and therefore
scales with the same exponent $\epsilon_{1}(h_{3})$ as the field independent one.
%
\section{Opening of a gap in a  longitudinal periodic field}
\label{sec:gap-longitudinal}
\begin{figure}[ht]
\centerline{\epsfig{file=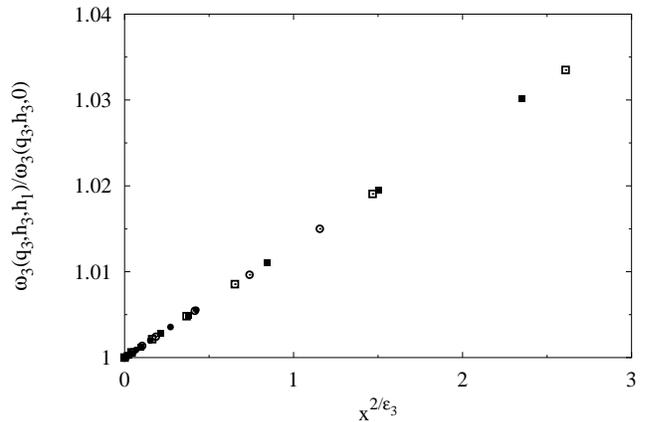,width=6.0cm,angle=-90}}
\caption{Finite-size scaling of the gap ratio~(\ref{eq:46}), 
  for $N=8,12,16,20$ and $q_{3}(M=1/4)=\pi/2$ 
  with $\epsilon_{3} = 0.81\ldots$}
\label{fig:fig_long}
\end{figure}

%
So far we have only considered the Hamiltonian~(\ref{eq:11}) with an
inhomogeneous field $h_{1}{\bf S}_{1}(\pi)$ transverse to the homogeneous field
$h_{3}{\bf S}_{3}(0)$. By means of the evolution equations~(\ref{eq:d2En})
and~(\ref{eq:dSmn}) we can also study the influence of a longitudinal periodic
field 
\begin{equation}
  \label{eq:32}
  {\bf H}(h_3,h_q) \equiv {\bf H}_{0} - 2h_{3} {\bf S_{3}}(0) + 
  2h_{q}{\bf \bar S}_{3}(q).
\end{equation}
The perturbation operator ${\bf \bar S}_{3}(q)\equiv[{\bf S}_{3}(q)+{\bf
  S}_{3}(-q)]/2$ commutes with the total spin operator $S_{T}^{3}$ and changes
the ground state momentum $p_{s}$ by $\pm q$. For this reason, all momentum
states with
\begin{equation}
  \label{eq:40}
  p_{k}=p_{s} \pm k q, \quad k=0,\pm 1,\pm 2,\ldots
\end{equation}
are coupled via the evolution equation. For example for $q=\pi/2$ there are 4
different momentum states with $p_{k}/\pi=\pm 1/2,0,1$,
which have to be taken into account. In general, the transition matrix elements
at $h_{q}=0$:
\begin{equation}
  \label{eq:41}
  T_{3}(h_{3},h_{q}=0) = 
     \langle p_{s}\pm q,S | {\bf S}_{3}(\pm q) |p_{s},S \rangle
\end{equation}
turn out to be finite, except for the case, where we meet a soft mode:
\begin{eqnarray}
  \label{eq:42}
  \omega_{3}(q,h_{3},h_{q}=0) =
   && E(p_{s}\!+\!q,S,h_{q}=0) -E(p_{s},S,h_{q}=0) \nonumber\\
  \stackrel{N\to\infty}{\longrightarrow}&& \frac{a_{3}(h_{3})}{N}
\end{eqnarray}
This happens if:
\begin{equation}
  \label{eq:43}
  q=q_{3}(M) =\pi(1-2M),
\end{equation}
e.g. a soft mode appears at $q=\pm\pi/2$ if $M=1/4$. At the soft
mode~(\ref{eq:43}) the transition matrix elements~(\ref{eq:41}) diverge:
\begin{equation}
  \label{eq:44}
  T_{3}(h_{3},0) 
    \stackrel{N\to\infty}{\longrightarrow} b_{3}(h_{3}) N^{\kappa_{3}(h_{3})}
\end{equation}
with an exponent $ \kappa_{3}(h_{3}) = 1 - \eta_{3}(M(h_{3}))/2$, given by the
$\eta_{3}(M)$-exponent, given in the introduction. From the evolution equations
with the initial conditions~(\ref{eq:42}) and~(\ref{eq:44}), we get in this case
a finite-size scaling behaviour of the gap ratio:
\begin{equation}
  \label{eq:46}
  \frac{\omega_{3}(q_{3},h_{3},h_{q})}{\omega_{3}(q_{3},h_{3},0)} 
   = 1 + e_{3}(x,h_{3}),
\end{equation}
with a scaling variable $x = Nh_{q}^{\epsilon_{3}(h_{3})}$, where
$\epsilon_{3}(h_{3}) = 1/[1+\kappa_{3}(h_{3})] $. The curve
$\epsilon_{3}(h_{3})$ is shown in Fig.~\ref{fig:eta}. Note, that
$\eta_{3}(M)=1/\eta_{1}(M)$, which means $\epsilon_{3}(0) = 2/3$, e.g. for
$M=1/4$, we have
\begin{equation}
  \label{eq:50}
  \epsilon_{3}(h_{3}(M=1/4)) = 0.81011\ldots.
\end{equation}
A test of the finite-size scaling behaviour~(\ref{eq:46}) for $q=\pi/2$ and
$M=1/4$ with the exponent~(\ref{eq:50}) is shown in Fig.~\ref{fig:fig_long}.
The small $x$-behaviour of the gap ratio is properly reproduced with
$x^{2/\epsilon_{3}}$ and compared with the prediction $h_{q}^{\epsilon_{3}}$,
where $\epsilon_{3}=\epsilon_{3}(h_{3}(M=1/4))$ is given by~(\ref{eq:50}). 
%
\subsection{The magnetization curve in a periodic field}\label{sec:mag}
\begin{figure}[ht]
\vspace{5cm}
\centerline{\hspace*{3.8cm}\epsfig{file=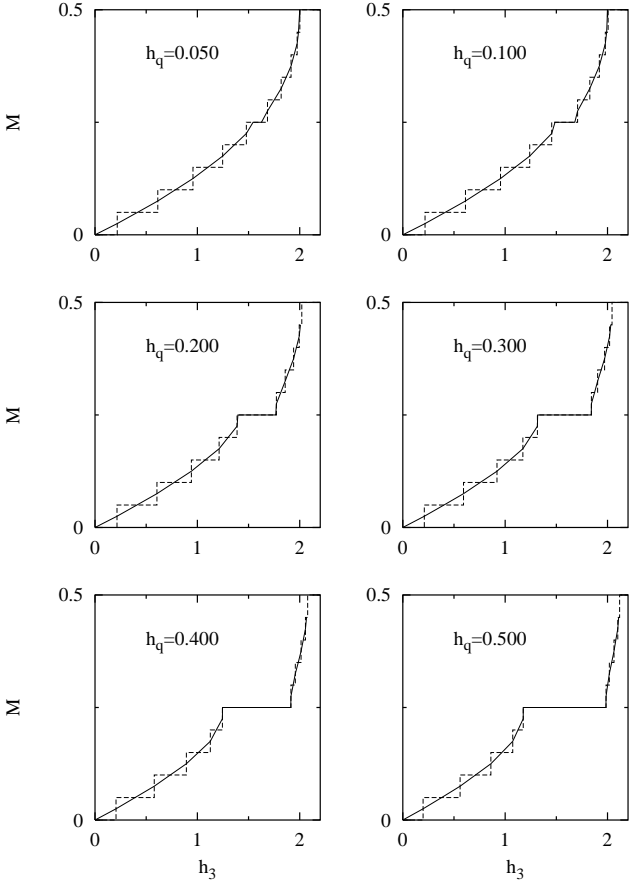,width=4.5cm}}
\caption{The evolution of a plateau in the magnetization curve at $M=1/4$,
  induced by an external field~(\ref{eq:32}) with period $q=\pi/2$. The
  magnetization curve is calculated from finite system ($N=20$) via midpoint
  magnetization \cite{BF64} in conjunction with a finite-size extrapolation of
  the plateau width from system sizes of $N=8,12,16,20$.}
\label{fig:fig_skal-plateau}
\end{figure}

\begin{figure}[ht]
\centerline{\epsfig{file=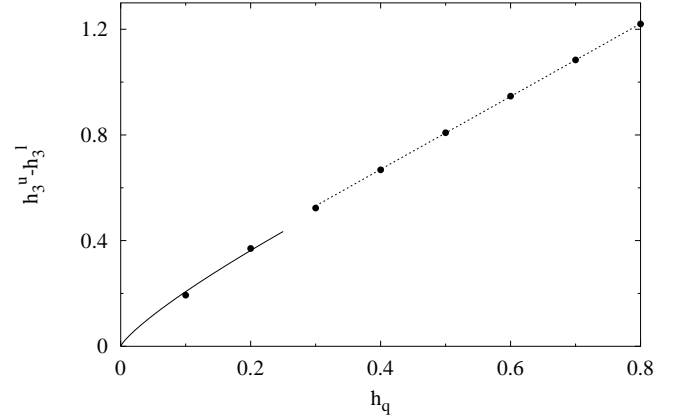,width=6.0cm,angle=-90}}
\caption{The evolution of the difference~(\ref{eq:51}) between the upper and
  lower critical field at the plateau $M=1/4$. The solid line shows a fit to the
  data for small values of the external periodic field $h_{q}$. The expected
  behaviour is $\propto h_{q}^{\epsilon_{3}}$, with $\epsilon_{3}=0.8101$ given
  by Eq.~(\ref{eq:50}).  The dashed line represents a linear fit for larger
  values of $h_{q}$.}
\label{fig:fig_bu-bl}
\end{figure}

%
Let us finally discuss the influence of the periodic perturbation
in~(\ref{eq:32}) on the magnetization curve $M=M(h_3)$. First of all one should
notice, that the opening of a gap for $h_{q}>0$ in the energy
differences~(\ref{eq:42}) does not imply a priori a plateau in the magnetization
curve. The criterium of a plateau with an upper and lower critical field
$h_3^u,h_3^l$ can be read from~(\ref{eq:33}) and~(\ref{eq:33b}):
\begin{eqnarray}
  \label{eq:51}
  2(h_3^u-h_3^l) &=& \lim_{N\to\infty} \left[
    E(p_{s}\!+\!\pi,S\!+\!1,h_{q})-2E(p_{s},S,h_{q}) \right. \nonumber \\
    &&\left. ~~~~~~~E(p_{s}\!+\!\pi,S\!-\!1,h_{q}) \right]. 
\end{eqnarray}
The emergence of the plateaus in the magnetization curve can be seen in
Fig.~\ref{fig:fig_skal-plateau}. A finite-size analysis shows that a non
vanishing difference~(\ref{eq:51}) remains in the thermodynamical limit. For
this analysis we have used the BST-Algorithm~\cite{BS64,CH93}. The
$h_{q}$-dependence of the plateau width is plotted in Fig.~\ref{fig:fig_bu-bl},
together with the predicted scaling behaviour $h_{q}^{\epsilon_{1}}$ for
$q=\pi/2$.
%
\section{Conclusions}
\label{sec:conclusions}
%
This paper is aimed to study the effect of a small periodic field on the
eigenvalue spectrum of the 1D spin-1/2 AFH model. We are interested in
particular in the opening of a gap in those situations, where the unperturbed
model is known to be critical. The critical exponents $\eta_{1}(M),\eta_{3}(M)$,
which govern the divergence in the transition matrix elements~(\ref{eq:7a}) and
~(\ref{eq:7b}) of the unperturbed model are known. Following conformal field
theory, they are related to the finite-size behaviour~(\ref{eq:3}) of certain
energy differences~(\ref{eq:2}) and~(\ref{eq:2b}), which can be computed on very
large systems by means of Bethe ansatz.

The evolution of the eigenvalue spectrum under the influence of perturbation of
strength $h_{q}$ is described by a system of differential
equations~(\ref{eq:d2En}) and~(\ref{eq:dSmn}), which has been shown to have
scaling solutions~(\ref{eq:29a}) and~(\ref{eq:29b}) in the scaling
limit~(\ref{eq:30}). The exponents $\epsilon$ and $\sigma$ in the scaling
solutions are uniquely determined by the corresponding $\eta$-exponents in the
unperturbed model. We have studied in detail the following types of
perturbations.
\begin{enumerate}
\item A transverse staggered field together with a homogeneous longitudinal
  field $h_{1}{\bf S}_{1}(\pi) + h_{3}{\bf S}_{3}(0)$. Both energy
  differences~(\ref{eq:2}) and~(\ref{eq:2b}) at the soft mode
  momenta~(\ref{eq:4}) were shown to evolve a gap with an exponent
  \begin{equation}
    \label{eq:39}
    \epsilon_{a}(h_{3}) = \frac{2}{4-\eta_{a}(M(h_{3}))},
  \end{equation}
with $a=1$ depending on the external homogeneous field $h_{3}$ with
magnetization $M(h_{3})$.
\item A longitudinal homogeneous and periodic field $2h_{3}{\bf
    S}_{3}(0)\\ +2h_{q}{\bf \bar S}_{3}(q)$. Such a perturbation
  creates a plateau in the magnetization curve $M=M(h_{3})$ at
  \begin{equation}
    \label{eq:45}
    M=\frac{1}{2}\left(1-\frac{q}{\pi}\right).
  \end{equation}
  In other words $q$ has to meet the soft mode momentum $q=q_{3}(M)=\pi(1-2M)$.
  The difference of the upper and lower critical field, which defines the width
  of the plateau, evolves with an exponent $\epsilon_{3}(h_{3})$, which is
  related to the corresponding $\eta_{3}$-exponent via~(\ref{eq:39}) for $a=3$.
\end{enumerate}
%
%
%
%

%
%

\end{document}